\documentclass[showpacs,natbib,preprintnumbers,amsmath,superscriptaddress,twocolumn,amssymb,prl]{revtex4}
\usepackage[dvips]{graphicx}
\usepackage{dcolumn}
\usepackage{bm}

\begin{document}
\title{Quantum depinning of the magnetic vortex core in micron-size permalloy disks}

\author{Ricardo Zarzuela}
\author{Sa\"{u}l V\'{e}lez}
\author{Joan Manel Hernandez}
\author{Javier Tejada}
 \email{jtejada@ubxlab.com}

\affiliation{Grup de Magnetisme, Departament de F\'{i}sica Fonamental, Universitat de
Barcelona, Barcelona 08028, Spain}

\author{Valentyn Novosad}
\affiliation{Materials Sciences Division, Argonne National Laboratory, Argonne, Illinois 60439, USA}


\date{\today}

\begin{abstract}
The vortex state, characterized by an in-plane closed flux domain structure and an out-of-plane magnetization at its centre (the vortex
core), is one of the magnetic equilibria of thin soft ferromagnetic micron-size dots. In the last two decades many groups have been
working on the dynamics of the magnetic moment in nanomagnetic materials at low temperatures, it giving rise to the observation of
quantum relaxations and quantum hysteresis cycles. For the first time, we report experimental evidence of quantum dynamics of the vortex
core of micron-size permalloy (Fe$_{19}$Ni$_{81}$) disks induced by the application of an in-plane magnetic field.  It is attributed to
the quantum tunneling of  the vortex core through pinning barriers, which are associated to structural defects in the dots, towards its
equilibrium position.
\end{abstract}

\maketitle

Equilibrium magnetic configurations of soft ferromagnetic materials can be essentially non-uniform.
It has been widely reported that, in the absence of applied magnetic field, micron-size disks of these materials exhibit
the vortex state (magnetic soliton) as the magnetic equilibrium of the system \cite{Cowburn,Shinjo,Novo1}.
Several technological and biomedical applications of the vortex state have been explored, such as nonvolatile memory devices \cite{Parkin},
biomolecular carriers \cite{Rozhkova} and targeted cancer-cell destruction \cite{Kim}.
The vortex state is characterized by a curling magnetization and an out-of-plane magnetic core, whose size is comparable to the
material's exchange length ($\sim 6$ nm). Striking feature of this magnetic equilibrium configuration is the unique role
played by the vortex core: despite the fact it does not have any significant effect on the static properties of the system
(it occupies far less than 0.01\% of the sample volume), the vortex core entirely governs the low frequency spin dynamics.
In particular, the excitation spectrum of these micron-size disks is characterized by the \textit{gyrotropic mode}, corresponding to
the spiral-like precessional motion of the vortex core as a whole \cite{Guslienko,Choe}. Its frequency belongs to the sub-GHz range,
and so it is intrinsically distinct from the spin wave modes. Furthermore, the chirality of this
spiral motion is determined by the vortex core polarization $p=\pm 1$ \cite{Choe}.

The vortex core is a suitable candidate to observe macroscopic quantum phenomena. Because of the strong
exchange interaction it behaves as an independent entity and, the vortex core being a nanoscopic object, it is feasible that it exhibits
quantum tunneling between classically stable configurations. The measurement of time relaxations of the magnetic moment is a simple experimental way to observe this phenomenon.
At finite temperature these relaxations may occur via thermal activation, whereas in the limit $T\rightarrow0$ these relaxations continue
independently of the temperature due to underbarrier quantum tunneling.
Macroscopic quantum tunneling (MQT) \cite{Caldeira-Leggett} determines that the relaxation rate decreases as $\exp(-S_{eff})$,
with $S_{eff}$ the action evaluated at the magnetic thermon (instanton), which takes into account the dynamics of the vortex core ruled by an
energy potential and the dissipation of the system at a given temperature. This behaviour has been widely observed in a large number of systems \cite{Eugene},
which include single domain particles \cite{Tejada,Vincent}, magnetic clusters \cite{Friedman-JM}, magnetic domain walls \cite{Domain Walls}, flux lines in type-II
superconductors \cite{Garcia-Santiago} and, very recently, Normal-Superconducting interfaces in type-I superconductors \cite{Velez}. All these
experimental evidences suggest that magnetic tunneling is a common phenomenon characterizing the low-temperature dynamics of magnetic
materials in the mesoscopic scale.

The application of an in-plane magnetic field yields the displacement of the vortex core perpendicularly to the field direction \cite{Shinjo} (see Fig. 1b).
It has been previously reported that the dynamics of the vortex core can be affected by the presence of structural defects in the sample
\cite{Shima,Compton}. The gyrotropic motion of vortices, being the softest dynamical mode, is the most likely
candidate for quantum relaxation. In the present letter we explore the magnetic irreversibility and the dynamics of
vortex cores in micron-size permalloy dot arrays at low temperatures by means of the application of an in-plane magnetic field. For the first time,
we report experimental evidence of the quantum depinning of magnetic vortex cores.

All measurements were performed in a commercial SQUID magnetometer capable to measure at temperatures down to 2K and to apply magnetic fields up to 5T. The system is equipped with a Continuous Low Temperature Control (CLTC) and an Enhanced Thermometry Control (ETC) and it showed thermal stability better than 0.01 K at all times in any isothermal measurement. We have studied an array of permalloy disks with diameter $2R=1.5\; \mu$m and thickness $L=95$ nm. Its surface density is $0.15\textrm{ dots}/\mu\textrm{m}^2$. Fig. 1a shows an AFM image of this array.  The array of permalloy disks was fabricated on a silicon wafer using optical lithography, and lift-off techniques: A single layer resist spin coating and highly directional electron-beam evaporation under UHV conditions were used to obtain circular dots with sharp edges. Optical lithography allows low cost patterning of a large area dot array in a single step. Consequently, identical properties of magnetic
material, such as grain size, distribution, and orientation, and film thickness may be obtained over the whole array. The magnetic film
was deposited on a water-cooled substrate from a permalloy (Fe$_{19}$Ni$_{81}$) target. The growth ratio was of ~1.5 \AA/s. The $95$ nm Py
film showed a switching field of about 4 Oe and appeared to be pretty isotropic (in-plane). Finally, the sample was prepared by stacking four $5\times5$ mm$^2$ of these arrays with parallel  sides and all magnetic measurements were performed using an in-plane configuration for the applied magnetic field. The sample was
studied in the range of temperatures $2-300$ K and under applied magnetic fields up to $2$ kOe.

\begin{figure}[htbp!]
\center
\includegraphics[scale=0.42]{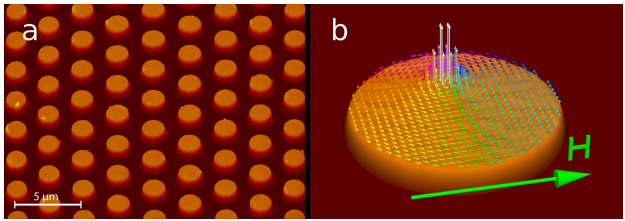}
\caption{\textbf{a}, AFM image of the array of permalloy disks studied in this paper.
\textbf{b}, Spin field of the vortex state in one of the permalloy disk considered in \textbf{a}. The vortex core is displaced
transversely to the direction of the applied field $H$.}
\end{figure}

%\section{Experimental results}

\begin{figure}[htbp!]
\center
\includegraphics[scale=0.29]{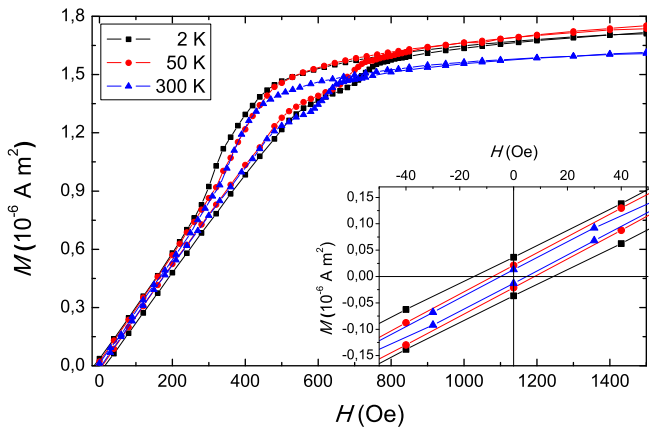}
\caption{$M(H)$ loops obtained at different temperatures (2, 50 and 300 K) for the range of positive applied magnetic
 fields. The inset shows their zoom in the field range from $H=-50$ Oe to $H=50$ Oe.}
\end{figure}

Fig. 2 shows the $M(H)$ hysteresis loop of the sample, in the range of positive applied magnetic fields, at different temperatures.
For the negative range, the cycles are antisymmetric.
The first magnetization curves have been omitted. Notice that these hysteresis loops correspond to the
single domain (SD)$\leftrightarrow$Vortex transitions \cite{Cowburn,Novo1}. As the temperature is lowered, the nucleation field $H_{n}$ decreases
and the annihilation field $H_{an}$ increases (as reported in ref. \cite{Mihajlovic}). For the range of temperatures explored,
the vortex linear regime in the ascending branch should extend from $-300$ Oe to $500$ Oe at least. Inset of Fig. 1 shows that the descending and ascending branches do not overlap at any temperature, and this extends over the whole linear regime. Furthermore, the remnant magnetization increases when $T$ decreases \cite{Shima}. In conclusion, the vortex linear regime exhibits magnetic irreversibility and it is temperature dependent.  Consequently, we proceed to explore the metastability of vortices by
means of i) ZFC-FC curves ($M_{ZFC}$ and $M_{FC}$) at different magnetic fields, and ii) isothermal measurements of the magnetization along the descending branch of the
hysteresis cycle ($M_{des}(H)$), from the SD state, at different $T$. In both i) and ii) the values of $T$ and $H$ at which the magnetization has been measured were the same.

\begin{figure}[htbp!]
\center
\includegraphics[scale=0.3]{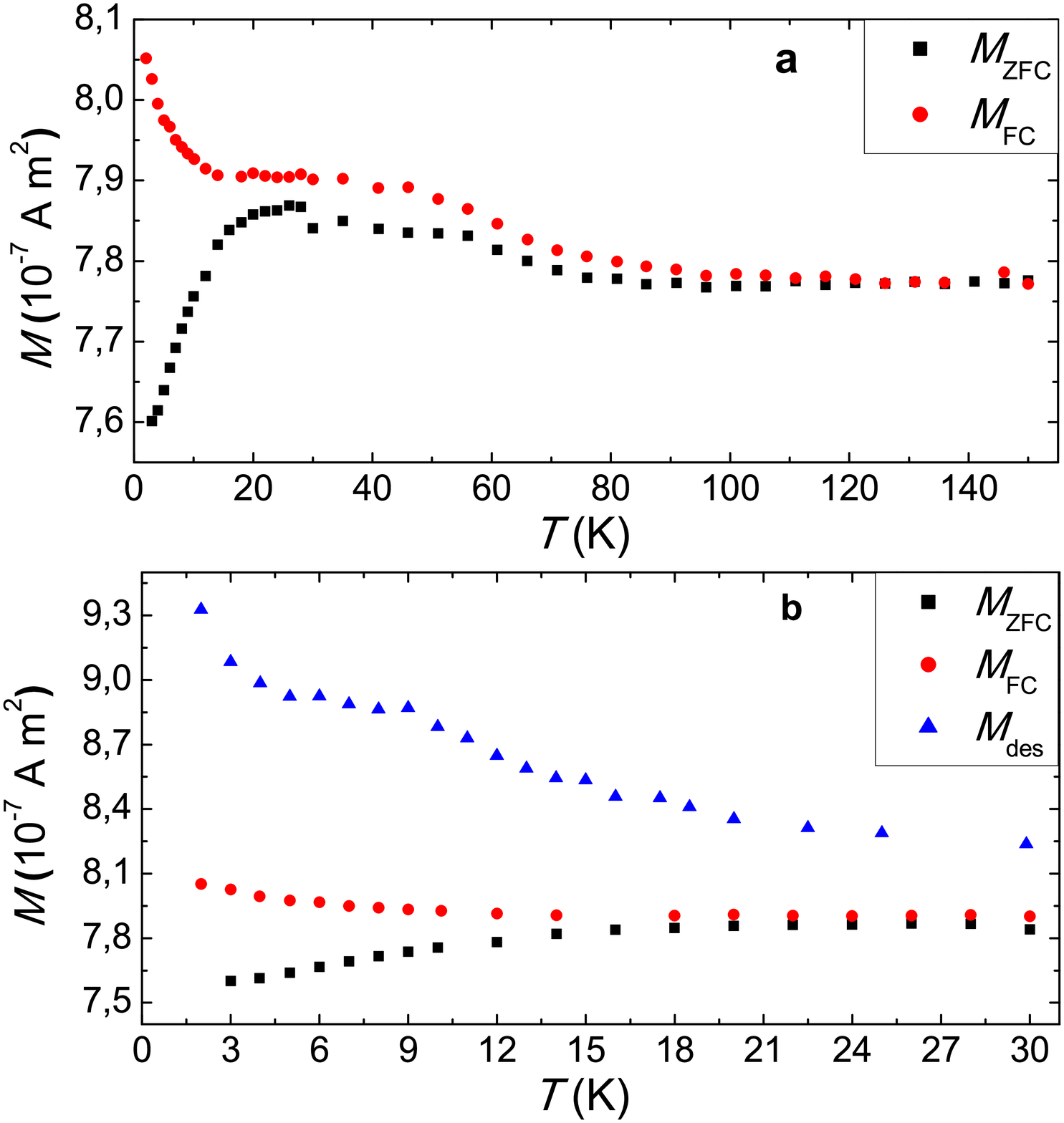}
\caption{\textbf{a}, Temperature dependence of $M_{ZFC}(300\textrm{ Oe})$ and $M_{FC}(300\textrm{ Oe})$ in the range $2-150$ K.
\textbf{b}, Plot of $M_{des}(300\textrm{ Oe})$, together with $M_{ZFC}(300\textrm{ Oe})$ and $M_{FC}(300\textrm{ Oe})$, in the range $2-30$ K. See text for details.}
\end{figure}

The ZFC process consists of first performing minor cycles around $H=0$ Oe at $T=150$ K (in order to get a zero magnetization state at $H=0$ Oe). Secondly, the sample was cooled down to $T=2$ K without applied magnetic field and, thirdly, a desired magnetic field, $H$, was applied. Then, the ZFC magnetization curve was measured from $2$ K to $150$ K. Sweeping back the temperature to 2 K we follow the FC curve. Fig. 3a shows the ZFC-FC curves obtained at $H=300$ Oe. The magnetization increases strongly from an initial value at 2 K to a maximum in the ballpark of $T\sim30$ K. Then it decreases smoothly and reaches a plateau. The dependence of the FC curve on $T$ is
similar to that of the ZFC case at high temperatures but with slightly higher values of $M$. In the ballpark of $T\sim20$ K, the magnetization
of the FC rises strongly, reaching its maximum value at $T=2$ K. This temperature dependence of both $M_{ZFC}(H)$ and $M_{FC}(H)$ is characteristic of the range of applied magnetic fields in the linear vortex regime. Additionally, isothermal magnetic measurements along
the descending branch of the hysteresis cycle ($M_{des}(H)$), from the SD state ($H=1$ kOe), have been measured at the same $T$ than the ZFC-FC curves. Fig. 3b shows $M_{des}(300\textrm{ Oe})$ obtained in the range $2-30$ K, together with $M_{ZFC}(300\textrm{ Oe})$ and $M_{FC}(300\textrm{ Oe})$. The values of $M_{des}(300\textrm{ Oe})$ decrease strongly when $T$ increases in the range $2-20$ K and above $T\sim30$ K tend smoothly to the FC curve (not shown). The divergence between the $	M_{ZFC}$, $M_{FC}$ and $M_{des}$ curves in the range $T=2-20$ K indicates the existence of a strong magnetic irreversibility in this region, and therefore we will focus on
this range from now on.

In order to confirm that the FC curve is the magnetic equilibria of the system, we performed two sets of measurements of
the isothermal time evolution of the magnetization, $M(T,t)$, when sweeping the temperature in increments of $1$ K per $30$ minutes,
i) from $15$ K to $2$ K  and ii) from $2$ K to $15$ K. The initial magnetic state for each set was prepared by means of the above ZFC process to
the desired temperature, followed by the application of a magnetic field $H=300$ Oe. In i) it is only observed magnetic relaxation of the sample
at $15$ and $14$ K, which quickly reaches a stable value corresponding to the FC one. From this point on, sweeping the temperature down to $2$ K
only leads to a variation of the magnetization of the sample following the values of the FC curve. ii) shows that, in the whole range of temperatures, the magnetization relaxes. The initial value of each relaxation follows the time evolution of the previous one. Moreover, the amount of relaxed magnetization is approximately the same for $T=2-9$ K and it
decreases progressively for $10-15$ K with magnetization values tending to the FC ones.

%\begin{figure}[htbp!]
%\center
%\includegraphics[scale=0.28]{Fig4}
%\caption{Isothermal time evolution of the magnetization of the sample, after application of $H=300$ Oe from the ZFC state,
%for different sweeps of $T$ in time steps of 30 minutes for each Kelvin. \textbf{a}, Sweeping from $15$ K to $2$ K.
%\textbf{b}, Sweeping from $2$ K to $15$ K.}
%\end{figure}

We explored the metastability of the system more deeply by performing relaxation measurements in the vortex linear regime. The amount of magnetization available
for relaxation is $M(0) - M_{eq}$, where $M(0)$ is the initial magnetization value and $M_{eq}$ corresponds to the equilibrium magnetization.
On account of this (see Fig. 3b), we will focus our study on the relaxation measurements of vortices from the metastable states of the descending branch.
Fig. 4a shows the normalized irreversible magnetization (left term of equation (1)) vs. $\ln t$ curves measured for two different applied fields ($H=0$ and $300$ Oe) at the same temperature
($T=2$ K). Only below $T\sim15$ K the magnetization of the sample fits very well a logarithmic time dependence. In this range of temperatures,
the magnetic viscosity $S(T)$ of the sample can be calculated by means of the theoretical formula \cite{Eugene}
\begin{equation}
\label{eq1}
\frac{M(t)-M_{eq}}{M(0)-M_{eq}}=1-S(T)\ln t
\end{equation}
Fig. 4b shows the viscosity, as a function of $T$, for two different magnetic fields. In both curves we observe a plateau at
low temperatures and, what is more important, it does not extrapolate to zero in the limit $T\rightarrow0$. In the ballpark of $T\sim7$ K,
the viscosity increases up to the temperature $\sim 11$ K, from which it decreases. Relaxation
measurements of the ZFC state with different applied magnetic fields were also performed, obtaining similar results for the viscosity.

\begin{figure}[htbp!]
\center
\includegraphics[scale=0.28]{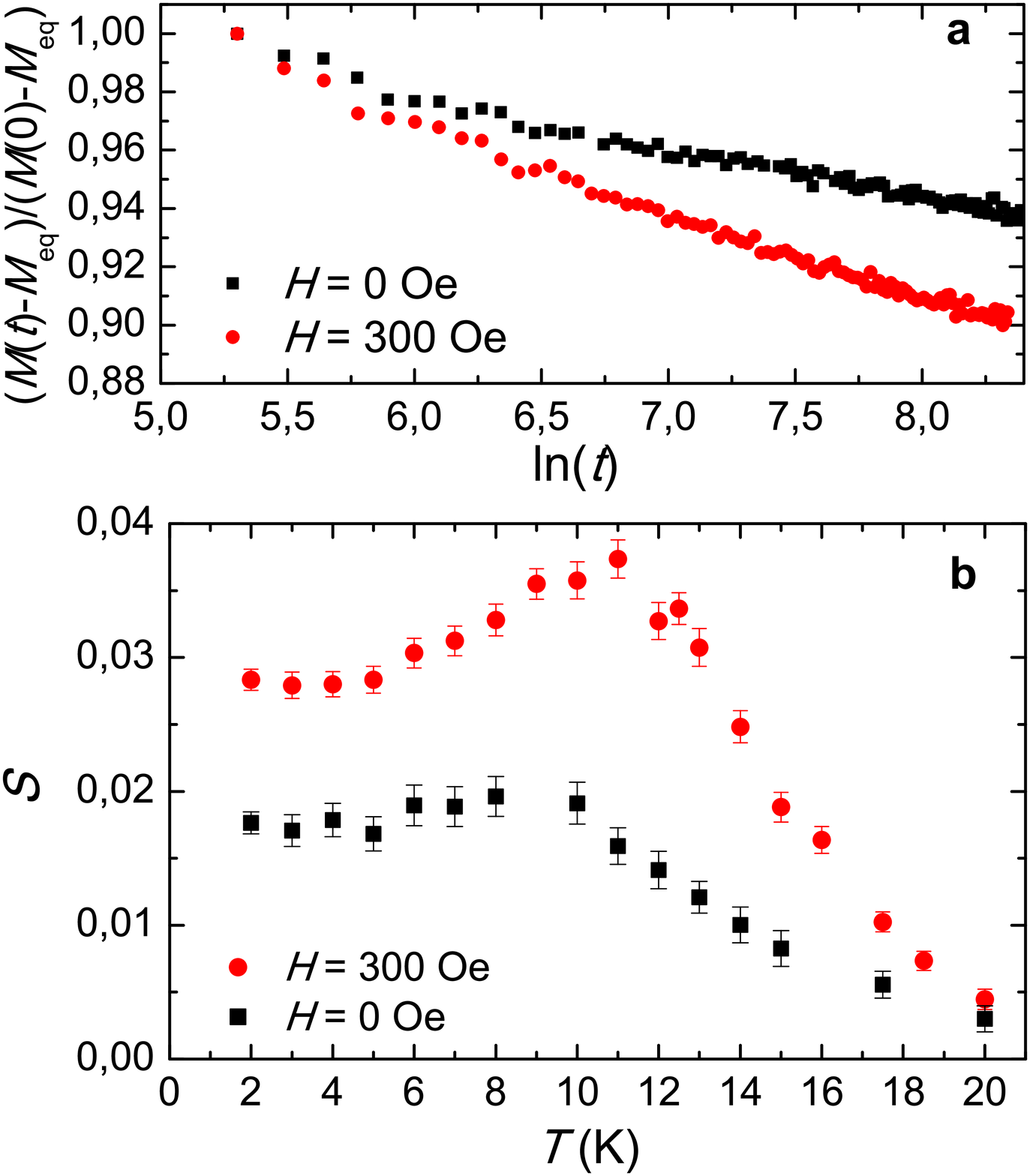}
\caption{\textbf{a}, Normalized irreversible magnetization vs. $\ln t$ curves measured at $T=2$ K for $H=0$ Oe and $300$ Oe.
\textbf{b}, Temperature dependence of the magnetic viscosity $S(T)$ at $H=0$ Oe and $300$ Oe. See text for details.}
\end{figure}

%\section{Discussion}
A logarithmic time dependence of the magnetization in relaxation measurements indicates the existence of a broad distribution of energy
barriers $U$ in our system. Classically, these energy barriers can be overcome by thermal activation, whose probability is proportional to the
Arrhenius factor $\exp(-U/T)$. The so-called Blocking temperature, $T_{B}$, sets apart both reversible ($T>T_{B}$) and irreversible ($T<T_{B}$) regimes
when the sample is externally perturbed. Despite the slight differences between $M_{ZFC}$ and $M_{FC}$ and between $M_{des}$ and $M_{FC}$
at high temperatures, the strong divergence of magnetization observed in these curves suggests that the Blocking
temperature should be below $T\sim 20$ K.

Conventionally, the Blocking temperature for weakly interacting systems can be estimated as the temperature at which the magnetic
viscosity reaches its maximum \cite{Eugene}. From our data we estimate that $T_{B}\sim 11$ K, which is in good agreement with the gradual
loss of logarithmic time dependence of our relaxation measurements at $T\gtrsim15$ K. Notice that thermal activation of energy
barriers dies out in the limit $T\rightarrow0$. Therefore, our observation that magnetic viscosity $S(T)$ tends to a finite value
different from zero as $T\rightarrow0$ indicates that relaxations are non-thermal in this regime, i.e., transitions
from metastable states are due to underbarrier quantum tunneling. This interpretation is upheld by the $M$ vs. $T\ln (t/\tau_{0})$ graphic
representation \cite{Vincent}. The time $\tau_{0}$ is the so called characteristic time attempt of the system and we have estimated its value
to be $\tau_{0}\sim10^{-11}$ s, so that all magnetic relaxation curves only scale for temperatures above $T\sim9$ K (see Fig. 5).
This loss of scaling corresponds to the quantum regime case and is independent of the energy barrier distribution.

\begin{figure}[htbp!]
\center
\includegraphics[scale=0.28]{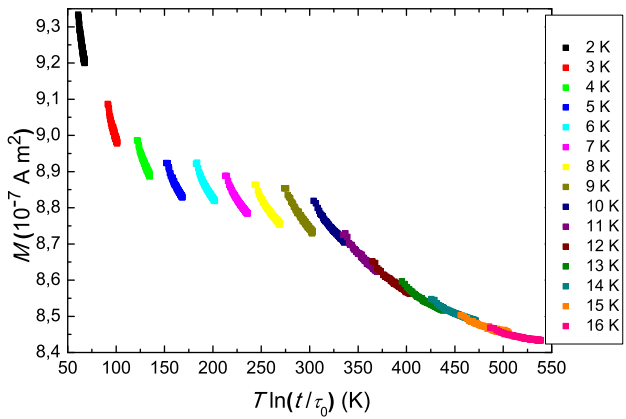}
\caption{Magnetization vs. $T\ln t$ curve measured at $H=300$ Oe. Above $T\sim9$ K, we verify the scaling $M=M(T\ln(t/\tau_{0}))$, which corresponds to the case of thermal relaxation. Below $T\sim9$ K we find a breakdown of this scaling, which corresponds to the quantum regime case.}
\end{figure}

The onset of irreversibility appears sweeping the external magnetic field for both the ZFC curve and the descending branch.
The effect of this field is to move the vortex core along the disk surface and the observed irreversibility should come from this movement.
Recent experimental data report the existence of some sort of structural defects in the disks \cite{Compton}, which could be a feasible origin of the
energy barriers responsible for the magnetic dynamics of the system. In the light of this, we consider these defects are capable of pinning
the vortex core. Therefore, the relaxation of the sample could be
understood as simply the dynamics of the vortex core when escaping from the pinning centers and overcoming the corresponding energy
barriers towards the equilibrium. Note that lower temperatures are probing lower energy barriers because of the $\exp(-U/T)$
dependence of the probability of thermal activation.

Within the framework of the rigid model of the shifted vortex, the vortex core is described as a zero-dimensional object
whose dynamics is ruled by Thiele's equation. The corresponding Langrangian is given by $\mathcal{L}=G y\dot{x}-W(\mathbf{r})$,
where $\mathbf{r}=(x,y)$ are the coordinates of the vortex core in the $XY$ plane, $G$ is the
modulus of its gyrovector and $W(\mathbf{r})$ is the total magnetic energy of the system. To incorporate the smallest pinning barriers into
the model we treat the vortex core as a stack of pancake vortices, one in each atomic layer.
This pancake structure shows a finite rigidity in the vertical dimension, which means that these layers interact elastically among
them. We consider that just a small vertical segment of the vortex core (equivalently, a small amount of these vortex layers) takes part in the tunneling
of the vortex core through the pinning barrier, whose length is $l<<L$. In order to
model all this we consider the vortex core as a flexible line that goes predominantly along the $\hat{z}$ direction, so that
$\mathbf{r}=\mathbf{r}(z,t)$ is a field depending on the vertical coordinate of the vortex core, $z$. The whole
magnetic energy (including the elastic and the pinning potential) is described via a biparametric quartic potential given by
$W(\mathbf{r})=-\mu h x+\frac{1}{2}\kappa (x^2+y^2)-\frac{1}{4}\beta x^4+\frac{\lambda}{2}\left(\frac{\partial \mathbf{r}}{\partial z}\right)^2$\cite{Zarzuela},
where $\mu$ and $h$ are the magnetic moment of the dot, respectively the modulus of external magnetic field (which is applied in
the $\hat{y}$ direction), $\lambda$ is the elastic coefficient of the pancake structure and $\kappa$ and $\beta$ are the parameters of the potential
energy. Assuming a second order transition for thermal to quantum relaxation, the obtained expressions for the crossover temperature $T_{c}$
and for the height of the barrier $U$ in absence of applied magnetic field ($h=0$) are the following \cite{Zarzuela}
\begin{equation}
k_{B} T_{c}\simeq\sqrt{5}\frac{\hbar \kappa}{2\pi G},\qquad U=\frac{\kappa^{2}}{4\beta}
\end{equation}
where the modulus of the gyrovector is given by the formula $G=2\pi (+1) l M_{s}/\gamma$, so that it is related to the tunneling vertical segment.

Comparison of the theoretical model with the experimental results leads to the determination of the parameters $(\kappa,\beta)$: first of all, notice
that $l$ cannot be smaller than the material's exchange length because, otherwise, the deformation of the vortex core line would be energetically
unfavorable to the system. The same happens if $l$ is much bigger than this exchange length. So it should be $l\sim l_{ex}=\sqrt{2A/\mu_{0}M_s^2}\simeq6$ nm, where $A=1.3\cdot10^{-11}$ J/m is the exchange constant and
$M_s=7.5\cdot10^{5}$ A/m is the saturation magnetization of permalloy. The value of the
modulus of the gyrovector is $G=2\pi (+1) l M_{s}/\gamma=1.62\cdot10^{-13}\textrm{ Ns/m}$ $(\gamma=1.76\cdot10^{11}\; \textrm{(Ts)}^{-1})$.
Experimentally, we have $T_{c}\sim 9$ K for the $H=0$ Oe case too, from which we deduce the value $\kappa\sim0.5 \textrm{ J/m}^2$. On the other hand,
for a measurable tunneling rate $S_{eff}$ should be in the ballpark of $30$. As $S_{eff}=c \kappa G/2\sqrt{2}\hbar\beta$ with $c$ being
a numerical factor of order unity resulting from the integration, we have the following estimate of the coefficient $\beta$,
$\beta\sim\kappa G/60\sqrt{2}\hbar=9.8\cdot10^{18}\textrm{ J/m}^{4}$. Finally, from these values of the parameters of
the pinning potential we can estimate the width of the energy barrier, which is given by the expression $L_{B}=\sqrt{2\kappa/\beta}
\sim0.3\textrm{ nm}$. Furthermore, we can also estimate the order of magnitude of the height of the barrier (mean value), $U\sim 250$ K,
which is in good agreement with the value given by the Blocking temperature. These estimates are feasible values because pinning happens at the
atomic level.

In conclusion, the non-thermal dynamics of magnetic vortices in micron-size permalloy disks is reported. It is attributed to the quantum depinning
of vortex cores through the structural defects of the sample, in steps about $0.3$ nm.

%\section{Acknowledgements}

R.Z. and S.V. acknowledge financial support from the Ministerio de Ciencia e Innovaci\'{o}n de Espa\~na.
J.T. acknowledges financial suport from ICREA Academia. The work at the University of Barcelona was funded by the
Spanish Government Project No. MAT2008-04535. The work at Argonne National Laboratory, including the use of facility at
the Center for Nanoscale Materials (CNM), was supported by UChicago Argonne, LLC, Operator of Argonne National Laboratory
("Argonne"). Argonne, a U.S. Department of Energy Office of Science Laboratory, is operated under Contract No. DE-AC02-06CH11357.

\end{document}